# Second and Third Harmonic Generation in Metal-Based Nanostructures

M. Scalora[1], M. A. Vincenti[2], D. de Ceglia[2], V. Roppo[3], M. Centini[4], N. Akozbek[2], M. J. Bloemer[1]

[1] *Charles M. Bowden Research Center AMSRD-AMR-WS-ST, RDECOM, Redstone Arsenal, Alabama 35898-5000, USA*

[2] *AEgis Technologies Group, 631 Discovery Dr., Huntsville, AL 35806*

[3] *Universitat Politècnica de Catalunya, Departament de Física i Eng. Nuclear, Colom 11, E-08222 Terrassa, Spain*

[4] *Dipartimento di Energetica, University of Rome La Sapienza, Via Scarpa 16, Rome Italy*

**Abstract**

We present a new theoretical approach to the study of second and third harmonic generation from metallic nanostructures and nanocavities filled with a nonlinear material, in the ultrashort pulse regime. We model the metal as a two-component medium, using the hydrodynamic model to describe free electrons, and Lorentz oscillators to account for core electron contributions to both the linear dielectric constant and to harmonic generation. The active nonlinear medium that may fill a metallic nanocavity, or be positioned between metallic layers in a stack, is also modeled using Lorentz oscillators and surface phenomena due to symmetry breaking are taken into account. We study the effects of incident TE- and TM-polarized fields and show that a simple re-examination of the basic equations reveals additional, exploitable dynamical features of nonlinear frequency conversion in plasmonic nanostructures.

**INTRODUCTION**

Interest in nonlinear frequency conversion in metals and semiconductors alike arches back to the beginning of nonlinear optics [1-15]. Recent research in linear plasmonic phenomena like sub-wavelength resolution [16] and enhanced transmission [17] has focused renewed attention on the origins of harmonic generation in metamaterials [18-23], metallic substrates with empty holes [24-27], holes filled with GaAs [28], resonant, sub-wavelength nanocavities [29], and layered metal-dielectric photonic and gap structures [30, 31]. These studies have shown that generation and enhancement of harmonic frequencies are possible in a variety of conditions and circumstances.

In the case of harmonic generation in metals it is notoriously difficult to reconcile quantitative and qualitative aspects of theory and experiments simultaneously. Usually, experimental results can be explained qualitatively by separating the nonlinear contributions into surface and volume sources, and by assigning to them suitable weights [35-39]. Our aim here is to study the dynamics in the ultrashort pulse regime, with an eye towards achieving as much qualitative and quantitative agreement as possible between theory and experiments, without imposing any separation between surface and volume sources. We treat free electrons using the hydrodynamic model [3, 40-42], make no a priori assumptions about charge or current distributions, and include Coulomb, Lorentz, convective, electron gas pressure, and linear and nonlinear contributions to the linear dielectric constant of the metal arising from bound (or valence) electrons. It has been shown that contributions to second harmonic (SH) generation from bound charges can be significant [10]. Free and bound electrons act in similar ways by displaying surface and volume sources, so that the general form of the nonlinear source may be specified in terms of a complex dielectric function defined at the fundamental and the SH frequencies [7, 11]. Bound electrons contribute to the linear dielectric constant of typical metals (interband transitions) at near-IR wavelengths for gold [43] and copper, with more pronounced effects in the visible and UV ranges [11, 44]. Even silver [45] departs from a simple Drude description at near-IR wavelengths. Its linear dielectric function is adequately described by a combined Drude-Lorentz model that contains a mix of free and bound electrons having one or more resonances at UV wavelengths [46-48]. That is:

$$\varepsilon(\omega) = 1 - \frac{\tilde{\omega}_{pf}^2}{\omega^2 + i\tilde{\gamma}_f \omega} - \frac{\tilde{\omega}_{pb}^2}{\omega^2 - \tilde{\omega}_{0,b}^2 + i\tilde{\gamma}_b \omega} \quad . \quad (1)$$

All parameters are scaled in units of $\mu m^{-1}$. $\tilde{\omega}_{p,f}, \tilde{\gamma}_f$ are the plasma frequency and damping coefficient for free electrons; $\tilde{\omega}_{p,b}, \tilde{\gamma}_b, \tilde{\omega}_{0,b}$ are the plasma frequency, damping coefficient and resonance frequency for bound electrons. By choosing $(\tilde{\omega}_{p,f}, \tilde{\gamma}_f) = (0.0573, 6.965)$ and $(\tilde{\omega}_{p,b}, \tilde{\gamma}_b, \tilde{\omega}_{0,b}) = (0.526, 3.96, 3.1)$

Eq.(1) becomes a fairly accurate representation of the dielectric function of silver down to approximately 400nm. In Fig.(1) we compare the data found in reference [45] with Eq.(1). The solid, black curve that runs below all others is the Drude portion of the right hand side of Eq.(1). The figure shows that it is not possible to fit any two points on the data curve using only a Drude function and a single plasma frequency, highlighting the importance of bound electrons. In this vein, in reference [30] the dynamics of the fundamental (800nm) and SH (400nm) fields was modeled using free electrons only. That kind of approach forces the use of two distinct, free-electron plasma frequencies for each field, even at wavelengths where the impact of core electrons is considerable. This modeling practice leads to the outright neglect of the dynamics of core electrons, and to the improper use of the linear data as if it originated only from free electrons. Instead, the mere introduction of bound electrons can sensibly recalibrate the linear dielectric function (phase) and its slope (group/energy velocity), which for computational purposes is akin to modifying effective plasma frequency, electron mass, and density [46]. Later in the examples we will see that small changes in the effective electron mass and density in conduction and valence bands can lead to notable qualitative and quantitative differences in the results.

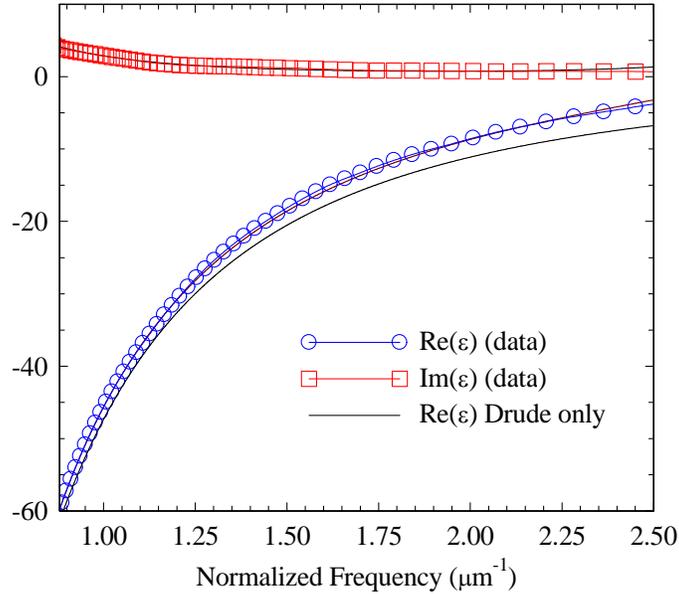

**Fig.1:** Markers: Real and imaginary parts of the dielectric constant of Ag in the 1064-400nm range. The solid curves that retrace the data correspond to Eq.(1). The lower solid curve that runs below all others is only the Drude portion of Eq.(1).

In contexts similar to those of reference [28], where a metallic nanocavity is filled with GaAs, there is a tendency to focus only on nonlinear restoring forces, to always neglect intrinsically nonlinear magnetic forces that drive all bound electrons, and even neglect harmonic generation arising from the metal itself. Indeed, while magnetic forces in bound electrons may be several orders of magnitude smaller than nonlinear restoring forces, they are always present nonetheless and in fact play a catalytic role by activating new interaction channels. The nonlinear frequency conversion of GaAs-filled holes on a gold substrate has been investigated at 3μm [28], with reported TE-polarized SH (1.5 μm) conversion efficiencies comparable to a standard quasi-phase-matched Lithium Niobate sample of similar thickness. In reference [29] it was reported that it is possible to calibrate the width of even a single aperture carved on a silver substrate of specified thickness in order to achieve enhanced transmission, field localization, and as a consequence, enhanced second harmonic generation that is strongly correlated to transmission maxima. The subwavelength cavity thus designed is capable of localizing and enhancing the incident fields by several orders of magnitude, with effects such as energy velocity reduction down to values less than $c/100$. It is therefore possible to amplify the nonlinear response of a nanocavity filled with a nonlinear material far more than reported in reference [28], for both TE- and TM-polarized harmonics, provided the cavity is aptly designed. These results will be reported separately.

**METALS**

Even if metals do not have intrinsic quadratic nonlinear terms, many early works consistently reported on SHG based on the mere presence of the Lorentz force induced by the incident magnetic field. Using a classical oscillator electron model, SH source terms consisting of a magnetic dipole due to the Lorentz force and an electric quadrupole-like contribution emerge also in centrosymmetric media [4-7]. Subsequent experimental work confirmed the existence of at least two SH source terms [8, 9]:

volume and surface contributions may be excited by a polarization normal or parallel to the plane of incidence, respectively. Even though this separation may be practical, it is generally recognized that it may not be possible to fully decompose nonlinear sources as surface and volume terms due to the presence of spatial derivatives on the bulk polarization [49]. These terms introduce some arbitrariness in the effective surface and volume coefficients, making a clear distinction between these two types of sources almost impossible to carry out [36].

The situation is more ambiguous for metal layers only a few tens of nanometers thick, where field penetration and localization occur inside the metal itself [30]. One might inquire about the relative importance of convective versus Lorentz or Coulomb terms [19], or how the presence of sharp corners in metallic nanocavities [29] or nanotubes [51] changes the relative contributions with respect to each other; or what role bound charges and electron gas pressure terms play. What follows is an attempt to address these issues and to highlight as much as possible the most salient dynamical aspects of the interaction of free and bound charges in the ultrashort pulse regime.

It is well-known that SHG can be enhanced by coupling with surface plasmons [15, 40]: the signal generated by reflection of the incident beam from a metal surface has important contributions from currents stimulated near the surface. In an effort to consider all the forces that act on the electrons as fully as possible in the ultrafast regime we have developed a detailed analysis of SHG and third harmonic generation (THG) from metallic surfaces and nanocavities that may contain a nonlinear material. The metal is composed of free electrons that occupy the conduction band (typically one electron per atom; for Ag the uppermost level is the $5s^1$) and electrons that fill the valence band (for Ag the uppermost filled valence level is the $4d^{10}$ orbital, with 10 available electrons). Free electrons are described by the hydrodynamic model [3, 40-42, 52], while ordinary Lorentz oscillators are used to describe bound charges. Although this is a simplified picture of metals, in what follows we derive and integrate new equations of motion that couple free and bound electrons in the metal to bound electrons in materials like GaAs, GaP, or LiNbO$_3$ in turn also described as a set of nonlinear Lorentz oscillators. All electrons are assumed to be under the influence of electric and magnetic forces, so that the dynamics that ensues in the metal and the dielectric contains surface and volume contributions simultaneously.

**FREE ELECTRONS**

An equation that describes free electrons inside the metal may be written as follows [3, 40]:

$$m^* \frac{d\mathbf{v}}{dt} + \gamma m^* \mathbf{v} = e\mathbf{E} + \frac{e}{c}\mathbf{v} \times \mathbf{H} - \frac{\nabla p}{n}; \qquad (2)$$

$m^*$ is the effective mass of conduction electrons and $n$ is their density; $\mathbf{v}$ is the electron velocity; $\mathbf{E}$ and $\mathbf{H}(=\mathbf{B})$ are electric and magnetic fields, respectively; $p$ is the pressure. The full temporal derivative of the velocity in Eq.(2) can be written as:

$$\frac{d\mathbf{v}}{dt} = \frac{\partial \mathbf{v}}{\partial t} + (\mathbf{v} \bullet \nabla)\mathbf{v}, \qquad (3)$$

so that Eq.(2) becomes:

$$\frac{\partial \mathbf{v}}{\partial t} + (\mathbf{v} \bullet \nabla)\mathbf{v} + \gamma \mathbf{v} = \frac{e}{m^*}\mathbf{E} + \frac{e}{m^*c}\mathbf{v} \times \mathbf{H} - \frac{\nabla p}{nm^*} \qquad . \qquad (4)$$

Identifying the current density with $\mathbf{J} = ne\mathbf{v}$ makes it possible to rewrite Eq.(4) as:

$$\frac{\partial \mathbf{J}}{\partial t} - \frac{\dot{n}}{n}\mathbf{J} + \mathbf{J} \bullet \nabla \left(\frac{\mathbf{J}}{ne}\right) + \gamma \mathbf{J} = \frac{ne^2}{m^*}\mathbf{E} + \frac{e}{m^*c}\mathbf{J} \times \mathbf{H} - \frac{\nabla p}{nm^*} \qquad . \qquad (5)$$

After defining $\dot{\mathbf{P}}_j = \mathbf{J}$, Eq.(5) becomes:

$$\ddot{\mathbf{P}}_j - \frac{\dot{n}}{n}\dot{\mathbf{P}}_j + (\dot{\mathbf{P}}_j \bullet \nabla)\left(\frac{\dot{\mathbf{P}}_j}{ne}\right) + \gamma \dot{\mathbf{P}}_j = \frac{ne^2}{m^*}\mathbf{E} + \frac{e}{m^*c}\dot{\mathbf{P}}_j \times \mathbf{H} - \frac{e\nabla p}{m^*} \qquad . \qquad (6)$$

For free electrons the continuity equation $\dot{n}(\mathbf{r},t) = -\frac{1}{e}\nabla \bullet \dot{\mathbf{P}}_j$ supplements the equations of motion, and may be integrated directly to yield:

$$n(\mathbf{r},t) = n_0 - \frac{1}{e}\nabla \bullet \mathbf{P}_j \quad , \tag{7}$$

where $n_0$ is the background, equilibrium charge density in the absence of any applied fields. In what follows our treatment departs from the usual procedure followed in the hydrodynamic model [3, 19, 40]. Assuming $\dot{n} << n$, the ratio $\dot{n}/n$ may be expanded in powers of $1/(n_0 e)$ to obtain:

$$\frac{\dot{n}}{n} = -\frac{1}{n_0 e}\nabla \bullet \dot{\mathbf{P}}\left(1 - \frac{1}{n_0 e}\nabla \bullet \mathbf{P}\right)^{-1} \sim -\frac{\nabla \bullet \dot{\mathbf{P}}}{en_0} - \frac{1}{n_0^2 e^2}(\nabla \bullet \dot{\mathbf{P}})(\nabla \bullet \mathbf{P}) + \mathcal{G}\left(\frac{1}{n_0^3 e^3}\right) + \dots \quad . \tag{8}$$

Substituting Eq.(8) back into Eq.(6) and neglecting terms of order $(1/n_0 e)^2$ and higher we get:

$$\ddot{\mathbf{P}}_j + \gamma \dot{\mathbf{P}}_j = \frac{n_0 e^2}{m^*}\mathbf{E} - \frac{e}{m^*}\mathbf{E}(\nabla \bullet \mathbf{P}_j) + \frac{e}{m^* c}\dot{\mathbf{P}}_j \times \mathbf{H} - \frac{1}{n_0 e}\left[(\nabla \bullet \dot{\mathbf{P}}_j)\dot{\mathbf{P}}_j + (\dot{\mathbf{P}}_j \bullet \nabla)\dot{\mathbf{P}}_j\right] - \frac{e\nabla p}{m^*} \quad . \tag{9}$$

In all the calculations that we performed for typical silver-based nanostructures ($n_0 \sim 5.8 \times 10^{22}/cm^3$) we consistently find $|\delta n| = |-\nabla \bullet \mathbf{P}_j/e| \sim 10^{13} - 10^{16}/cm^3$. The lower bound is typical of uniform metal layers, while the upper bound is characteristic of resonant subwavelength metallic slits and nanocavities [29].

The specific impact of pressure is seldom considered in the dynamics [40], but it is instructive to make a few observations in its regard. Pressure may be treated classically by assuming that electrons form an ideal gas, i.e. $p = n K_B T$ [53]. $K_B$ is the Boltzman constant. The divergence of $p$ in Eq.(9) becomes the divergence of $n$, which in turn may be related to the macroscopic polarization as follows:

$$-\frac{e\nabla p}{m^*} = -\frac{e}{m^*}K_B T \nabla\left(n_0 - \frac{1}{e}\nabla \bullet \mathbf{P}_j\right) = \frac{K_B T}{m^*}\nabla(\nabla \bullet \mathbf{P}_j) \quad . \tag{10}$$

It is interesting and equally instructive to also look at a quantum model of the pressure [54]. The quantum pressure is typically described as $p = p_0 (n/n_0)^\gamma$, where $\gamma = (D+2)/D$ and D is the dimensionality of the problem [54, 55]. For D=3, we have $p = p_0 (n/n_0)^{5/3}$, where $p_0 = n_0 E_F$, $E_F$ is the Fermi energy, and $n_0$ is again the equilibrium charge density. The leading pressure terms are:

$$-\frac{e\nabla p}{m^*} = -\frac{ep_0}{m^* n_0^{5/3}} \frac{5}{3} n^{2/3} \nabla n = -\frac{5}{3} \frac{en_0 E_F}{m^* n_0^{5/3}} n^{2/3} \nabla n \approx \frac{5}{3} \frac{E_F}{m^*} \nabla(\nabla \bullet \mathbf{P}_j) - \frac{10}{9} \frac{E_F}{m^*} \frac{1}{n_0 e} (\nabla \bullet \mathbf{P}_j) \nabla(\nabla \bullet \mathbf{P}_j). \quad (11)$$

As already shown elsewhere using a two-fluid quantum model [54], and as Eq.(11) plainly suggests, the quantum model intrinsically contains a first order classical, ideal electron gas contribution (if we equate the Fermi energy and $K_B T$ in Eq.(10) ) and a nonlinear quantum correction of lower order. It is easier to see the impact of pressure if we scale the equations with respect to dimensionless time, longitudinal and transverse coordinates, $\tau = ct/\lambda_0$, $\xi = z/\lambda_0$, $\tilde{y} = y/\lambda_0$, respectively, where $\lambda_0 = 1\mu m$ is chosen as the reference wavelength. As a result of this scaling Eq.(9) becomes:

$$\ddot{\mathbf{P}}_j + \tilde{\gamma}\dot{\mathbf{P}}_j = \frac{n_0 e^2}{m^*}\left(\frac{\lambda_0}{c}\right)^2 \mathbf{E} - \frac{e\lambda_0}{m^* c^2}\mathbf{E}(\nabla \bullet \mathbf{P}_j) + \frac{e\lambda_0}{m^* c^2}\dot{\mathbf{P}}_j \times \mathbf{H} - \frac{1}{n_0 e \lambda_0}\left[(\nabla \bullet \dot{\mathbf{P}}_j)\dot{\mathbf{P}}_j + (\dot{\mathbf{P}}_j \bullet \nabla)\dot{\mathbf{P}}_j\right]$$
$$+ \frac{5}{3}\frac{E_F}{m^* c^2}\nabla(\nabla \bullet \mathbf{P}_j) - \frac{10}{9}\frac{E_F}{m^* c^2}\frac{1}{n_0 e \lambda_0}(\nabla \bullet \mathbf{P}_j)\nabla(\nabla \bullet \mathbf{P}_j) \quad , (12)$$

In addition to the magnetic Lorentz force, $(e\lambda_0 / m^* c^2)\dot{\mathbf{P}}_j \times \mathbf{H}$, we have an explicit quadrupole-like [1] Coulomb term that arises from the continuity equation, $-(e\lambda_0 / m^* c^2)\mathbf{E}(\nabla \bullet \mathbf{P}_j)$, convective terms proportional to $\left[(\nabla \bullet \dot{\mathbf{P}}_j)\dot{\mathbf{P}}_j + (\dot{\mathbf{P}}_j \bullet \nabla)\dot{\mathbf{P}}_j\right]$, and linear and nonlinear pressure terms proportional to $\nabla(\nabla \bullet \mathbf{P}_j)$ and $(\nabla \bullet \mathbf{P}_j)\nabla(\nabla \bullet \mathbf{P}_j)$, respectively. For silver, the Fermi velocity $v_F \sim 10^8$ cm/sec and $(E_F / m^* c^2) \sim 10^{-5}$, which is then combined with $1/(n_0 e \lambda_0) \sim 10^{-10}$ (cgs, silver). If for the moment we neglect all nonlinear contributions, Eq.(12) becomes:

$$\ddot{\mathbf{P}}_j + \tilde{\gamma}\dot{\mathbf{P}}_j = \frac{n_0 e^2}{m^*}\left(\frac{\lambda_0}{c}\right)^2 \mathbf{E} + \frac{5}{3}\frac{E_F}{m^* c^2}\nabla(\nabla \bullet \mathbf{P}_j). \quad (13)$$

Expanding the terms on right hand side of Eq.(13) shows that the pressure couples orthogonal, free electron polarization states and introduces a dynamical anisotropy. More generally, pressure could directly impact the linear dielectric function of the metal near its walls, should the fields become strongly confined and/or their derivatives be large enough [56] (i.e. near sharp edges, corners, or in resonant subwavelength cavities) to introduce large, evanescent k-vectors. The same is true for the nonlinear term: its magnitude could perturb Coulomb, Lorentz, or convective terms at high-enough intensity and/or if large-enough k-vectors were excited. With these issues in mind, some simple considerations may be made about nonlinear sources derivable from Eq.(12). Assuming the incident pump is undepleted and time harmonic, lowest order terms may be collected as follows:

$$\mathbf{P}_{NL,free}^{SH}(2\omega) \approx -\frac{e\lambda_0}{m^*c^2}\mathbf{E}_\omega \nabla \bullet \left(\chi_{free}(\omega)\mathbf{E}_\omega\right) - \frac{e\lambda_0}{m^*c^2}\chi_{free}(\omega)i\beta\mathbf{E}_\omega \times \mathbf{H}_\omega$$
$$-\frac{\beta^2\chi_{free}(\omega)}{n_0 e\lambda_0}\left[\mathbf{E}_\omega \nabla \bullet \left(\chi_{free}(\omega)\mathbf{E}_\omega\right) + \left(\mathbf{E}_\omega \bullet \nabla\right)\chi_{free}(\omega)\mathbf{E}_\omega\right] - \frac{10}{9}\frac{E_F}{m^*c^2}\frac{1}{n_0 e\lambda_0}\left[\nabla \bullet \left(\chi_{free}(\omega)\mathbf{E}_\omega\right)\right]\nabla\left[\nabla \bullet \left(\chi_{free}(\omega)\mathbf{E}_\omega\right)\right]$$
, (14)

where $\beta = 2\pi\omega/\omega_0$ and $\chi_{free}(\omega)$ is the free-electron portion of the dielectric function. A similar equation may be written for the third harmonic polarization. Should the incident signal be a short pulse [25, 30], material dispersion is included to all orders in Eq.(12) along with all boundary conditions [38, 39]. The form of the spatial derivatives in Eq.(14) suggests that convective and quantum terms have properties similar to the Coulomb term. However, this could change for thin layers, subwavelength nanocavities, or for wavelengths in the near IR, visible and UV ranges, where the fields can penetrate and become localized inside the metal [30, 31]. For this reason our only reference point will be Eq.(12).

For localized ultrashort pulses harmonic generation occurs regardless of angle of incidence or polarization state. Additional dynamical features may be ascertained by decomposing Coulomb and Lorentz forces explicitly into all their harmonic components. The geometry of our system is shown in Fig.2. Two fields of orthogonal polarizations and arbitrary amplitudes are incident on a structure whose details are unspecified. The interaction takes place on the y-z plane and the fields are independent of x. Up to TH frequency, TE- and TM-polarized fields (according to Fig.(2)) are as follows, respectively:

$$\mathbf{E}_{TE} = \mathbf{i}E_{TEx} = \mathbf{i}\left(E_{TEx}^{\omega}e^{-i\omega t} + \left(E_{TEx}^{\omega}\right)^{*}e^{i\omega t} + E_{TEx}^{2\omega}e^{-2i\omega t} + \left(E_{TEx}^{2\omega}\right)^{*}e^{2i\omega t} + E_{TEx}^{3\omega}e^{-3i\omega t} + \left(E_{TEx}^{3\omega}\right)^{*}e^{3i\omega t}\right)$$

$$\mathbf{H}_{TE} = \begin{pmatrix} \mathbf{j}H_{TEy} \\ +\mathbf{k}H_{TEz} \end{pmatrix} = \mathbf{j}\left(H_{TEy}^{\omega}e^{-i\omega t} + \left(H_{TEy}^{\omega}\right)^{*}e^{i\omega t} + H_{TEy}^{2\omega}e^{-2i\omega t} + \left(H_{TEy}^{2\omega}\right)^{*}e^{2i\omega t} + H_{TEy}^{3\omega}e^{-3i\omega t} + \left(H_{TEy}^{3\omega}\right)^{*}e^{3i\omega t}\right) \; ; (15)$$

$$+\mathbf{k}\left(H_{TEz}^{\omega}e^{-i\omega t} + \left(H_{TEz}^{\omega}\right)^{*}e^{i\omega t} + H_{TEz}^{2\omega}e^{-2i\omega t} + \left(H_{TEz}^{2\omega}\right)^{*}e^{2i\omega t} + H_{TEz}^{3\omega}e^{-3i\omega t} + \left(H_{TEz}^{3\omega}\right)^{*}e^{3i\omega t}\right)$$

$$\mathbf{E}_{TM} = \begin{pmatrix} \mathbf{j}E_{TMy} \\ +\mathbf{k}E_{TMz} \end{pmatrix} = \mathbf{j}\left(E_{TMy}^{\omega}e^{-i\omega t} + \left(E_{TMy}^{\omega}\right)^{*}e^{i\omega t} + E_{TMy}^{2\omega}e^{-2i\omega t} + \left(E_{TMy}^{2\omega}\right)^{*}e^{2i\omega t} + E_{TMy}^{3\omega}e^{-3i\omega t} + \left(E_{TMy}^{3\omega}\right)^{*}e^{3i\omega t}\right)$$

$$+\mathbf{k}\left(E_{TMz}^{\omega}e^{-i\omega t} + \left(E_{TMz}^{\omega}\right)^{*}e^{i\omega t} + E_{TMz}^{2\omega}e^{-2i\omega t} + \left(E_{TMz}^{2\omega}\right)^{*}e^{2i\omega t} + E_{TMz}^{3\omega}e^{-3i\omega t} + \left(E_{TMz}^{3\omega}\right)^{*}e^{3i\omega t}\right) . (16)$$

$$\mathbf{H}_{TM} = \mathbf{i}H_{TMx} = \mathbf{i}\left(H_{TMx}^{\omega}e^{-i\omega t} + \left(H_{TMx}^{\omega}\right)^{*}e^{i\omega t} + H_{TMx}^{2\omega}e^{-2i\omega t} + \left(H_{TMx}^{2\omega}\right)^{*}e^{2i\omega t} + H_{TMx}^{3\omega}e^{-3i\omega t} + \left(H_{TMx}^{3\omega}\right)^{*}e^{3i\omega t}\right)$$

The field envelopes in Eqs.(15-16) are not assumed to be slowly varying. The extraction of carrier frequencies is done only as a matter of convenience. With reference to Fig.2, both TE- and TM-polarized fields have transverse components orthogonal to each other, but they should not be confused: one component points along $\hat{\mathbf{x}}$, the other along $\hat{\mathbf{y}}$. Eqs.(15) and (16) become:

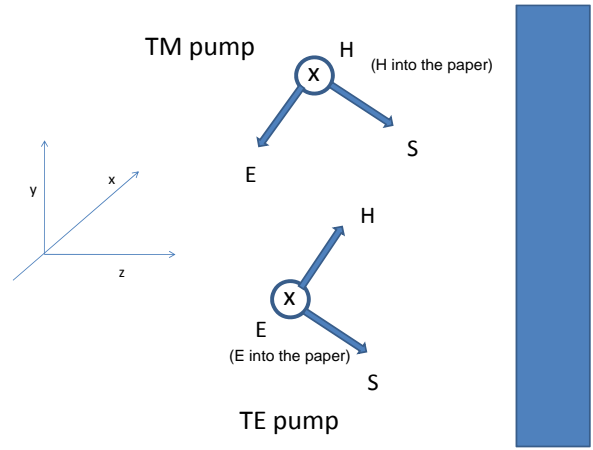

**Fig.2:** Incident TE and TM polarized fields. In all our cases the fields overlap, but appear spatially separated for clarity.

$$\mathbf{E} = \mathbf{i}E_{TEx} + \mathbf{j}E_{TMy} + \mathbf{k}E_{TMz} \qquad \mathbf{H} = \mathbf{i}H_{TMx} + \mathbf{j}H_{TEy} + \mathbf{k}H_{TEz} . \qquad (17)$$

In similar fashion the electric polarization has TE and TM components, and may be expressed as:

$$\mathbf{P} = (P_{TEx}\mathbf{i} + P_{TMy}\mathbf{j} + P_{TMz}\mathbf{k}). \tag{18}$$

An expansion of the magnetic Lorentz force using Eqs.(17-18) yields the following constituents:

$$\begin{aligned}\dot{\mathbf{P}} \times \mathbf{H} &= \left(\dot{P}_{TEx}\mathbf{i} + \dot{P}_{TMy}\mathbf{j} + \dot{P}_{TMz}\mathbf{k}\right) \times \left(H_{TMx}\mathbf{i} + H_{TEy}\mathbf{j} + H_{TEz}\mathbf{k}\right) \\ &= \left(\dot{P}_{TMy}H_{TEz} - \dot{P}_{TMz}H_{TEy}\right)\mathbf{i} + \left(\dot{P}_{TMz}H_{TMx} - \dot{P}_{TEx}H_{TEz}\right)\mathbf{j} + \left(\dot{P}_{TEx}H_{TEy} - \dot{P}_{TMy}H_{TMx}\right)\mathbf{k}\end{aligned} \tag{19}$$

These equations show that even a normally incident, TE-polarized field generates a non-zero, TM-polarized harmonic component via the term $\left(\dot{P}_{TEx}H_{TEy}\right)\mathbf{k}$, thus opening a possible catalytic interaction channel. At oblique incidence the term $\left(-\dot{P}_{TEx}H_{TEz}\right)\mathbf{j}$ also provides non-zero gain for a TM-polarized harmonic signal. The Coulomb term may also be decomposed as follows:

$$\left(\nabla \bullet \mathbf{P}_j\right)\mathbf{E} = \left(\frac{\partial P_{TMz}}{\partial z} + \frac{\partial P_{TMy}}{\partial y}\right)\left(\mathbf{i}E_{TEx} + \mathbf{j}E_{TMy} + \mathbf{k}E_{TMz}\right), \tag{20}$$

where $\partial P_{TEx}/\partial x = 0$. Therefore an incident beam with a mixed polarization state activates all possible conversion channels. Simultaneously TE/TM-polarized incident fields lead to efficiencies that may be different and much enhanced compared to pumping with either TE- or TM-polarized light only.

**BOUND ELECTRONS**

Bound electrons differ from free electrons in at least two ways: (1) they may be under the action of linear and nonlinear restoring forces; (2) the average local charge density remains constant in time, as electrons are not free to leave their atomic sites. The method that we present here to treat Coulomb and Lorentz forces was developed in reference [4], and may be combined with nonlinear restoring forces, i.e. $\chi^{(2)}$, $\chi^{(3)}$ or higher order processes, to generalize the nonlinear response of dielectrics or semiconductors to include surface phenomena dynamically. For example, the ability to describe the simultaneous excitation of linear and nonlinear plasmonic phenomena is important to model semiconductor nanocavities and slits in the UV range, where the dielectric function may be negative [50].

Neglecting for the moment nonlinear restoring forces, Newton' second law for one species of core electrons leads to an effective polarization equation for bound charges that reads as follows:

$$\ddot{\mathbf{P}}_b + \gamma_b \dot{\mathbf{P}}_b + \omega_{0,b}^2 \mathbf{P}_b = \frac{n_{0,b} e^2}{m_b^*} \mathbf{E} + \frac{e}{m_b^* c} \dot{\mathbf{P}}_b \times \mathbf{H} \quad . \tag{21}$$

Here $\mathbf{P}_b = n_{0,b} e \mathbf{r}_b$ is the polarization; $\mathbf{r}_b$ is the electron's position relative to an equilibrium origin; $n_{0,b}$ is the constant density; $m_b^*$ is the electron's effective mass in the valence band; $\dot{\mathbf{P}}_b = n_{0,b} e \dot{\mathbf{r}}_b$ is the bound current density. Up to the third harmonic frequency, the fields may be written as:

$$\begin{aligned}\mathbf{E} &= \left( \mathbf{E}_\omega e^{i(\mathbf{k} \cdot \mathbf{r}_b - \omega t)} + \mathbf{E}_\omega^* e^{-i(\mathbf{k} \cdot \mathbf{r}_b - \omega t)} + \mathbf{E}_{2\omega} e^{2i(\mathbf{k} \cdot \mathbf{r}_b - \omega t)} + \mathbf{E}_{2\omega}^* e^{-2i(\mathbf{k} \cdot \mathbf{r}_b - \omega t)} + \mathbf{E}_{3\omega} e^{3i(\mathbf{k} \cdot \mathbf{r}_b - \omega t)} + \mathbf{E}_{3\omega}^* e^{-3i(\mathbf{k} \cdot \mathbf{r}_b - \omega t)} \right) \\ \mathbf{H} &= \left( \mathbf{H}_\omega e^{i(\mathbf{k} \cdot \mathbf{r}_b - \omega t)} + \mathbf{H}_\omega^* e^{-i(\mathbf{k} \cdot \mathbf{r}_b - \omega t)} + \mathbf{H}_{2\omega} e^{2i(\mathbf{k} \cdot \mathbf{r}_b - \omega t)} + \mathbf{H}_{2\omega}^* e^{-2i(\mathbf{k} \cdot \mathbf{r}_b - \omega t)} + \mathbf{H}_{3\omega} e^{3i(\mathbf{k} \cdot \mathbf{r}_b - \omega t)} + \mathbf{H}_{3\omega}^* e^{-3i(\mathbf{k} \cdot \mathbf{r}_b - \omega t)} \right)\end{aligned} \tag{22}$$

For simplicity in what follows we assume incident plane waves, although the same arguments may be extended to more generic fields. Expanding the fields in powers of $\mathbf{k} \cdot \mathbf{r}_b$ we have [4]:

$$\mathbf{E} = \begin{pmatrix} \mathbf{E}_\omega e^{-i\omega t} \left( 1 + i\mathbf{k} \cdot \mathbf{r}_b + \frac{(i\mathbf{k} \cdot \mathbf{r}_b)^2}{2} + \ldots \right) + \mathbf{E}_\omega^* e^{i\omega t} \left( 1 - i\mathbf{k} \cdot \mathbf{r}_b + \frac{(-i\mathbf{k} \cdot \mathbf{r})^2}{2} + \ldots \right) \\ + \mathbf{E}_{2\omega} e^{-2i\omega t} \left( 1 + 2i\mathbf{k} \cdot \mathbf{r}_b + \frac{(2i\mathbf{k} \cdot \mathbf{r}_b)^2}{2} + \ldots \right) + \mathbf{E}_{2\omega}^* e^{2i\omega t} \left( 1 - 2i\mathbf{k} \cdot \mathbf{r}_b + \frac{(-2i\mathbf{k} \cdot \mathbf{r}_b)^2}{2} + \ldots \right) \\ + \mathbf{E}_{3\omega} e^{-3i\omega t} \left( 1 + 3i\mathbf{k} \cdot \mathbf{r}_b + \frac{(3i\mathbf{k} \cdot \mathbf{r}_b)^2}{2} + \ldots \right) + \mathbf{E}_{3\omega}^* e^{3i\omega t} \left( 1 - 3i\mathbf{k} \cdot \mathbf{r}_b + \frac{(-3i\mathbf{k} \cdot \mathbf{r}_b)^2}{2} + \ldots \right) \end{pmatrix}, \tag{23}$$

and similarly for the magnetic field. The solutions for the electron's position and its derivatives are:

$$\mathbf{r}_b = \mathbf{r}_\omega e^{-i\omega t} + \mathbf{r}_\omega^* e^{i\omega t} + \mathbf{r}_{2\omega} e^{-2i\omega t} + \mathbf{r}_{2\omega}^* e^{2i\omega t} + \mathbf{r}_{3\omega} e^{-3i\omega t} + \mathbf{r}_{3\omega}^* e^{3i\omega t}, \tag{24}$$

$$\dot{\mathbf{r}}_b = -i\omega \mathbf{r}_\omega e^{-i\omega t} + i\omega \mathbf{r}_\omega^* e^{i\omega t} - 2i\omega \mathbf{r}_{2\omega} e^{-2i\omega t} + 2i\omega \mathbf{r}_{2\omega}^* e^{2i\omega t} - 3i\omega \mathbf{r}_{3\omega} e^{-3i\omega t} + 3i\omega \mathbf{r}_{3\omega}^* e^{3i\omega t} \tag{25}$$

$$\ddot{\mathbf{r}}_b = -\omega^2 \mathbf{r}_\omega e^{-i\omega t} - \omega^2 \mathbf{r}_\omega^* e^{i\omega t} - 4\omega^2 \mathbf{r}_{2\omega} e^{-2i\omega t} - 4\omega^2 \mathbf{r}_{2\omega}^* e^{2i\omega t} - 9\omega^2 \mathbf{r}_{3\omega} e^{-3i\omega t} - 9\omega^2 \mathbf{r}_{3\omega}^* e^{3i\omega t}. \tag{26}$$

After substituting Eq.(22-26) into Eq.(21) the solutions may be written compactly as:

$$\mathbf{r}_\omega = \frac{e}{m(-\omega^2+\omega_0^2)} \begin{pmatrix} \mathbf{E}_\omega - i\mathbf{k}\cdot\mathbf{r}_{2\omega}\mathbf{E}_\omega^* \\ -2i\mathbf{k}\cdot\mathbf{r}_{3\omega}\mathbf{E}_{2\omega}^* + 2i\mathbf{k}\cdot\mathbf{r}_\omega^*\mathbf{E}_{2\omega} \\ +3i\mathbf{k}\cdot\mathbf{r}_{2\omega}^*\mathbf{E}_{3\omega} \end{pmatrix} + \frac{e}{mc(-\omega^2+\omega_0^2)} \begin{pmatrix} +i\omega\mathbf{r}_\omega^* \times \mathbf{H}_{2\omega} - 2i\omega\mathbf{r}_{2\omega}\times\mathbf{H}_\omega^* \\ +2i\omega\mathbf{r}_{2\omega}^*\times\mathbf{H}_{3\omega} - 3i\omega\mathbf{r}_{3\omega}\times\mathbf{H}_{2\omega}^* \\ -\omega\mathbf{r}_\omega\times\mathbf{H}_\omega^*\mathbf{k}\cdot\mathbf{r}_\omega - \omega\mathbf{r}_\omega^*\times\mathbf{H}_\omega\mathbf{k}\cdot\mathbf{r}_\omega + \omega\mathbf{r}_\omega\times\mathbf{H}_\omega\mathbf{k}\cdot\mathbf{r}_\omega^* \end{pmatrix}$$

$$\mathbf{r}_{2\omega} = \frac{e}{m(-4\omega^2+\omega_0^2)} \begin{pmatrix} +\mathbf{E}_{2\omega} \\ +i\mathbf{k}\cdot\mathbf{r}_\omega\mathbf{E}_\omega - i\mathbf{k}\cdot\mathbf{r}_{3\omega}\mathbf{E}_\omega^* \\ +3i\mathbf{k}\cdot\mathbf{r}_\omega^*\mathbf{E}_{3\omega} \end{pmatrix} + \frac{e}{mc(-4\omega^2+\omega_0^2)} \begin{pmatrix} -i\omega\mathbf{r}_\omega\times\mathbf{H}_\omega - 3i\omega\mathbf{r}_{3\omega}\times\mathbf{H}_\omega^* + i\omega\mathbf{r}_\omega^*\times\mathbf{H}_{3\omega} \\ -2\omega\mathbf{r}_{2\omega}\times\mathbf{H}_\omega^*\mathbf{k}\cdot\mathbf{r}_\omega - 2\omega\mathbf{r}_\omega\times\mathbf{H}_\omega^*\mathbf{k}\cdot\mathbf{r}_{2\omega} \\ -\omega\mathbf{r}_\omega^*\times\mathbf{H}_\omega\mathbf{k}\cdot\mathbf{r}_{2\omega} - 2\omega\mathbf{r}_\omega^*\times\mathbf{H}_{2\omega}\mathbf{k}\cdot\mathbf{r}_\omega \\ +2\omega\mathbf{r}_{2\omega}\times\mathbf{H}_\omega\mathbf{k}\cdot\mathbf{r}_\omega^* + 2\omega\mathbf{r}_\omega\times\mathbf{H}_{2\omega}\mathbf{k}\cdot\mathbf{r}_\omega^* \end{pmatrix}$$

$$\mathbf{r}_{3\omega} = \frac{e}{m(-9\omega^2+\omega_0^2)} \begin{pmatrix} +\mathbf{E}_{3\omega} \\ +i\mathbf{k}\cdot\mathbf{r}_{2\omega}\mathbf{E}_\omega + 2i\mathbf{k}\cdot\mathbf{r}_\omega\mathbf{E}_{2\omega} \end{pmatrix} + \frac{e}{mc(-9\omega^2+\omega_0^2)}\left(-2i\omega\mathbf{r}_{2\omega}\times\mathbf{H}_\omega - i\omega\mathbf{r}_\omega\times\mathbf{H}_{2\omega} + \omega\mathbf{r}_\omega\times\mathbf{H}_\omega\mathbf{k}\cdot\mathbf{r}_\omega\right)$$

(27)

In writing Eqs.(27) we have excluded higher order terms that couple magnetic and electric dipoles, terms like $\left(\omega\mathbf{r}_\omega\times\mathbf{H}_\omega\mathbf{k}\cdot\mathbf{r}_\omega\right)$ that already appear in Eq.(27) but that contain at least two harmonic fields. These terms could become important in solids at TW/cm$^2$ or in highly nonlinear plasmas [57, 58]. Of course, one should consider these and other terms if different pumping conditions are used, and matters should be evaluated on a case by case basis [59]. Finally, Eq.(27) may be simplified if we neglect higher order magnetic dipole-electric dipole terms, and if we identify:

$$\begin{aligned}
\mathbf{P}_\omega &= n_{0,b}e\mathbf{r}_\omega & \mathbf{P}_{2\omega} &= n_{0,b}e\mathbf{r}_{2\omega} & \mathbf{P}_{3\omega} &= n_{0,b}e\mathbf{r}_{3\omega} \\
\dot{\mathbf{P}}_\omega &\approx -i\omega n_{0,b}e\dot{\mathbf{r}}_\omega & \dot{\mathbf{P}}_{2\omega} &\approx -2i\omega n_{0,b}e\dot{\mathbf{r}}_{2\omega} & \dot{\mathbf{P}}_{3\omega} &\approx -3i\omega n_{0,b}e\dot{\mathbf{r}}_{3\omega} \\
i\mathbf{k}\cdot n_{0,b}e\mathbf{r}_\omega &\approx \nabla\cdot\mathbf{P}_\omega & 2i\mathbf{k}\cdot n_{0,b}e\mathbf{r}_{2\omega} &\approx \nabla\cdot\mathbf{P}_{2\omega} & 3i\mathbf{k}\cdot n_{0,b}e\mathbf{r}_{3\omega} &\approx \nabla\cdot\mathbf{P}_{3\omega}
\end{aligned} \quad (28)$$

Then, Eq.21 finally becomes:

$$\ddot{\mathbf{P}}_{b,\omega} + \tilde{\gamma}_b\dot{\mathbf{P}}_{b,\omega} + \tilde{\omega}_{0,b}^2\mathbf{P}_{b,\omega} \approx \frac{n_{0,b}e^2}{m_b^*}\left(\frac{\lambda_0}{c}\right)^2\mathbf{E}_\omega + \frac{e\lambda_0}{m_b^*c^2}\left(-\frac{1}{2}\mathbf{E}_\omega^*\nabla\cdot\mathbf{P}_{b,2\omega} + 2\mathbf{E}_{2\omega}\nabla\cdot\mathbf{P}_{b,\omega}^*\right) + \frac{e\lambda_0}{m_b^*c^2}\begin{pmatrix} +\dot{\mathbf{P}}_{b,\omega}^*\times\mathbf{H}_{2\omega} + \dot{\mathbf{P}}_{b,2\omega}\times\mathbf{H}_\omega^* \\ +\dot{\mathbf{P}}_{b,2\omega}^*\times\mathbf{H}_{3\omega} + \dot{\mathbf{P}}_{b,3\omega}\times\mathbf{H}_{2\omega}^* \end{pmatrix}$$

$$\ddot{\mathbf{P}}_{b,2\omega} + \tilde{\gamma}_b\dot{\mathbf{P}}_{b,2\omega} + \tilde{\omega}_{0,b}^2\mathbf{P}_{b,2\omega} = \frac{n_{0,b}e^2}{m_b^*}\left(\frac{\lambda_0}{c}\right)^2\mathbf{E}_{2\omega} + \frac{e\lambda_0}{m_b^*c^2}\begin{pmatrix} \mathbf{E}_\omega\nabla\cdot\mathbf{P}_{b,\omega} \\ -\frac{1}{3}\mathbf{E}_\omega^*\nabla\cdot\mathbf{P}_{b,3\omega} - 3\mathbf{E}_{3\omega}\nabla\cdot\mathbf{P}_{b,\omega}^* \end{pmatrix} + \frac{e\lambda_0}{m_b^*c^2}\begin{pmatrix} \dot{\mathbf{P}}_{b,\omega}\times\mathbf{H}_\omega \\ +\dot{\mathbf{P}}_{b,\omega}^*\times\mathbf{H}_{3\omega} + \dot{\mathbf{P}}_{b,3\omega}\times\mathbf{H}_\omega^* \end{pmatrix}$$

$$\ddot{\mathbf{P}}_{b,3\omega} + \tilde{\gamma}_b\dot{\mathbf{P}}_{b,3\omega} + \tilde{\omega}_{0,b}^2\mathbf{P}_{b,3\omega} = \frac{n_{0,b}e^2}{m_b^*}\left(\frac{\lambda_0}{c}\right)^2\mathbf{E}_{3\omega} + \frac{e\lambda_0}{m_b^*c^2}\left(\frac{1}{2}\mathbf{E}_\omega\nabla\cdot\mathbf{P}_{b,2\omega} + 2\mathbf{E}_{2\omega}\nabla\cdot\mathbf{P}_{b,\omega}\right) + \frac{e\lambda_0}{m_b^*c^2}\left(\dot{\mathbf{P}}_{b,2\omega}\times\mathbf{H}_\omega + \dot{\mathbf{P}}_{b,\omega}\times\mathbf{H}_{2\omega}\right)$$

. (29)

Expanding Eq.(12) into all its harmonics leads to equations similar to Eqs.(29). Eqs.(12) and (29) are a new set of coupled equations that describe free and bound charges that give rise to second and third harmonic generation in metallic nanostructures. Eqs.(29) are also applicable to dielectrics and semiconductors alike, and are valid for pump depletion and harmonic down-conversion, or cases where the harmonic fields may be more intense than the fundamental field. The introduction of nonlinear restoring forces generalizes Eq.(29) to describe nonlinear phenomena in semiconductor nanocavities. This may be achieved by adding nonlinear terms to the right hand sides of each of Eqs.(29), or by introducing a nonlinear polarization in the usual way:

$$\mathbf{P}_{NL} = \chi^{(2)}\mathbf{E}^2 + \chi^{(3)}\mathbf{E}^3 + ... \quad . \quad (30)$$

A comparison between Eqs.(12) and (29) shows that at this order of approximation the equations have similar form, even though the Coulomb terms have different origins. For free electrons the equation of continuity accounts for charge density variations in time. In the case of bound charges it is the spatial variation of the fields that leads to similar contributions. The Coulomb terms have opposite signs, negative for free electrons and positive for bound charges, with identifiable, effective charge distributions and currents that add to and interfere with their free electron counterparts. Given the similarities between Eqs.(12) and (29), it is reasonable to expect comparable qualitative responses from free and bound charges [7, 11]. We also note that it is possible to arrive at slightly different forms of Eqs.(29) that involve the fields explicitly rather than the polarization, as was done in reference [4].

We conclude this section with a few words about expectations from the model. What remains to be done to solve the general problem is to couple Eqs.(12) and (29) to Maxwell's equations. The complete set of coupled equations that results from expanding up to the third harmonic fields is solved using a time domain, split-step beam propagation method. The technique is outlined in reference [60], where it was developed for the wave equation to describe envelope functions that vary slowly in time only. However, since Maxwell's equations are first order in time, the method is easily extended by removing all approximations to take into account all orders of reflection and dispersion. The integration scheme is stable, and the evaluation of spatial derivatives across hard interfaces is straightforward.

As may be easily ascertained from Eqs.(12) and (29), the only free parameters that the model allows are the effective electron masses and number densities in conduction and valence bands. These parameters are important because they determine the quantitative aspect of conversion efficiencies. Although the values of electron mass and density are well-known, in practice there continues to be uncertainty and so the issue has been addressed often in the literature [61-64]. The actual values appear to be sensitive not only to the method of growth and layer thickness [64], but also depend on the natural porosity of the metal under consideration and the surface on which it is deposited [65, 66]. Reference [64] reports that using D.C. magnetron sputtering, layers of gold exhibited effective conduction electron masses that ranged from $0.48m_e$ to $1.14m_e$ for 9.6nm and 22nm layers, respectively. Similar results were found for silver, with deduced effective masses that varied from $1.06m_e$ to $1.99m_e$ for 10nm and 19nm layers, respectively. The situation improves somewhat for R.F. sputtering, but the effective masses still vary with layer thickness between 20% for silver (from $1.02m_e$ to $0.84m_e$) and 35% for gold (from $1.38m_e$ to $1.04m_e$). As a consequence of these relatively large fluctuations, it is reasonable to assume that the electron number density, plasma frequency, skin depth, and Fermi energy will also vary accordingly, thus allowing substantial flexibility in departing from tabulated values. Although the natural porosity of metals can be used to control and tune plasmonic behavior of gold across a large wavelength range [66], it can also manifest itself in a somewhat disquieting manner [65]. The experimental results reported in reference [65] indicated that a silver layer deposited on a glass substrate may be considerably more porous on the air side than on the glass side. By monitoring surface plasmons propagating on either side, it was determined that for an incident wavelength of 632nm the real part of the dielectric "constant" on the air side was approximately 25% smaller compared to the glass side, while the imaginary part could vary by as much as 60%. These differences are significant, persist for thick layers, and ultimately modify the quantitative aspects of linear propagation (actual reflection and transmission) and harmonic generation. Finally, there is the issue of the surface roughness. It is easy to imagine how in the linear regime roughness could act in ways similar to porosity, in that it could lead to changes of the effective mass, density, skin depth, and Fermi energy. In the nonlinear regime the enhancement of SHG due to surface roughness has already been discussed [67]: it takes place in a manner similar to what occurs in the case of surface enhanced Raman scattering [56], and as such it is also characterized by the formation of SH hot spots. Although under normal conditions all these issues may be neglected in favor of qualitative rather than quantitative agreement (e.g. uniform layers), the discussion above provides hints that sometimes the interpretations of linear and/or nonlinear plasmonic phenomena that revolve around subwavelength slits, holes, sharp edges, or extremely narrow, nanometer-wide plasmonic transmission resonances or band gaps may be more subtle and intricate than previously thought.

**THE IMPACT OF CONVECTION**

We now turn to the assessment of convective terms on harmonic generation. Fig.3 shows the calculated SHG efficiency upon reflection from a 125nm-thick silver film suspended in air of a TM-polarized, ~1GW/cm$^2$ pump pulse approximately 50fs in duration, tuned to 1064nm. We measure efficiency as the ratio of total energy converted/total initial pump energy after the pulse has passed, and show results with and without convective contributions. The detected radiation is also TM-polarized. In the case of a single, uniform thick layer (see caption of Fig.3 for the parameters used in the calculation) the quantitative effect of convection at 1064nm is evidently to increase conversion efficiency by a factor of ~2.5. We note that conversion efficiency quickly converges for pulses just a few optical cycles in duration, so that the same results are obtained for longer pulses. Even though in the case of Fig.3 conversion efficiency is enhanced, convective contributions are sensitive to the geometry of the structure and their impact cannot be generalized as simply at it might appear from Fig.3. As an example we calculated transmitted and reflected second harmonic conversion efficiencies from a transparent metal/dielectric multilayer stack similar to those studied in references [30] and [31]. The structure consists of a 5-period $Ta_2O_5$(125nm)/Ag(20nm) stack, with an incident, 1GW/cm$^2$, 60fs pulse tuned to 800nm. The stack and its transmission function are depicted in Fig.4. In Fig.5 we report the predicted transmitted and reflected conversion efficiencies for TM-incident/TM-detected second harmonic generation vs. incident angle, with and without convective terms. The model we use presently is much improved compared to the model that was employed in reference [30], where the nonlinear source was derived only from the Lorentz force. Also significant is the fact that in this model dielectric interfaces become active via the presence of the spatial derivatives of the polarization in Eqs.(29), that explicitly account for symmetry breaking contributions from all bound electron sources. In other words, this stack would still generate harmonics via Eq.(29) even if we turned off all the nonlinear terms in Eq.(12). Our calculations suggest that in this case convection tends to suppress and shift the peaks of conversion efficiency. With convection the reflected SH signal is smaller compared to its counterpart without convection by ~50%, and the peak is shifted toward larger angles. For the transmitted SH signal the tendency is more dramatic compared to the single layer: the maximum turns into a minimum. However, these trends could change for different layer thicknesses, and this example should suffice to highlight the sensitivity of the process to convective terms. The quantitative consequences of convection are far more dramatic if the stack of Fig.4 is pumped with a mixed TE/TM polarization state having equal TE- and TM- polarized incident field amplitudes and TE-polarized detected signal. The results are depicted in Fig.6. The figure shows that in addition to shifting the peaks to larger angles, the inclusion of convective terms can also reduce reflected and transmitted second harmonic conversion efficiencies in this case by a factor of 50. Usually neither a single TE- nor a single TM-polarized incident field can generate a TE-polarized SH signal without additional degrees of freedom and/or nonlinear sources. However, the dual-pumping mode opens up and couples all available interaction channels, so that TE- and TM-polarized SH and TH light may be produced more efficiently. These results show that in the general case the presence of convection can influence the dynamics and strongly impact the results quantitatively and qualitatively. Field penetration and localization inside the metal may accentuate its importance.

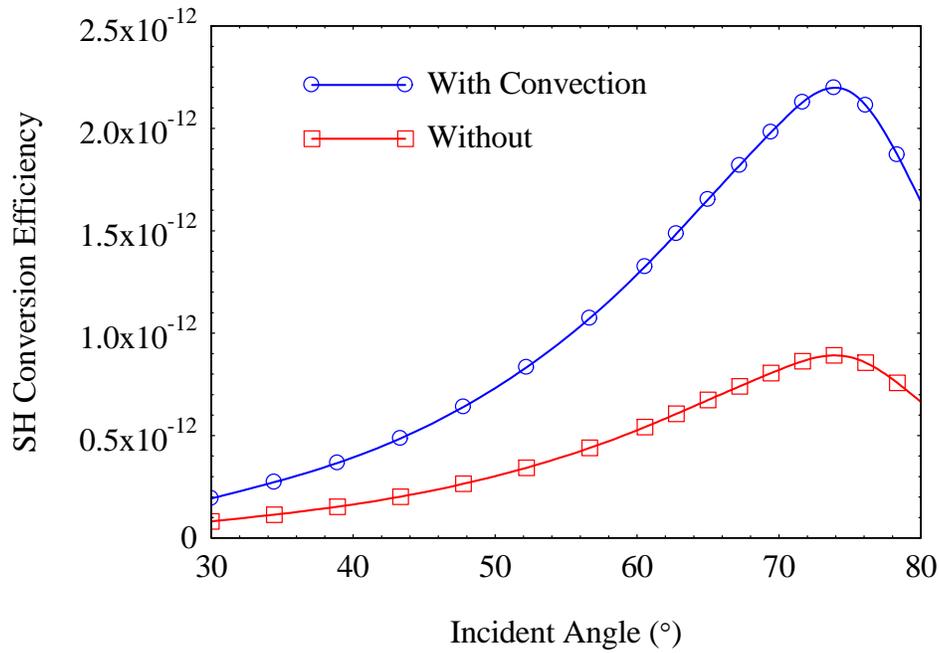

**Fig.3:** Conversion efficiency for TM-incident (1064nm)/TM-detected (632nm) light for a 125nm-thick silver layer. The effective mass and densities of both free and bound charges are taken to be $m^*=m_e$, and $n_0=5.8\times10^{22}/cm^3$, respectively.

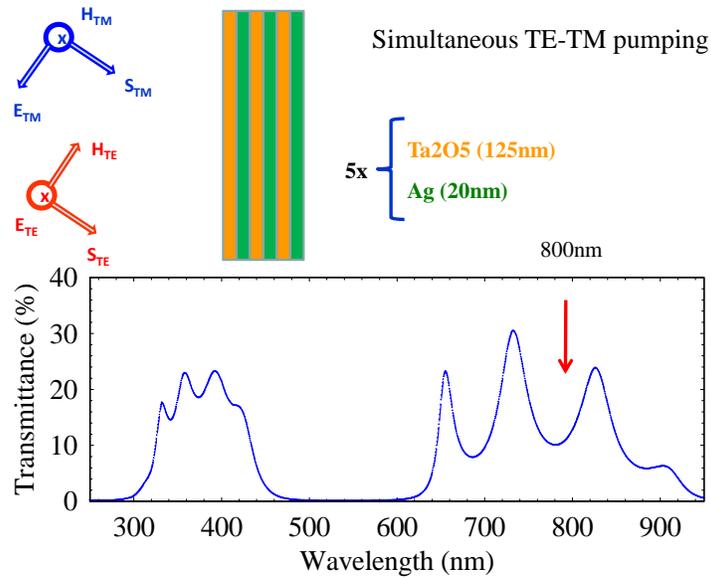

**Fig.4:** Transmission from a 5-period Ta2O5/Ag stack at normal incidence. **Inset**: picture of the incident pulse and the stack.

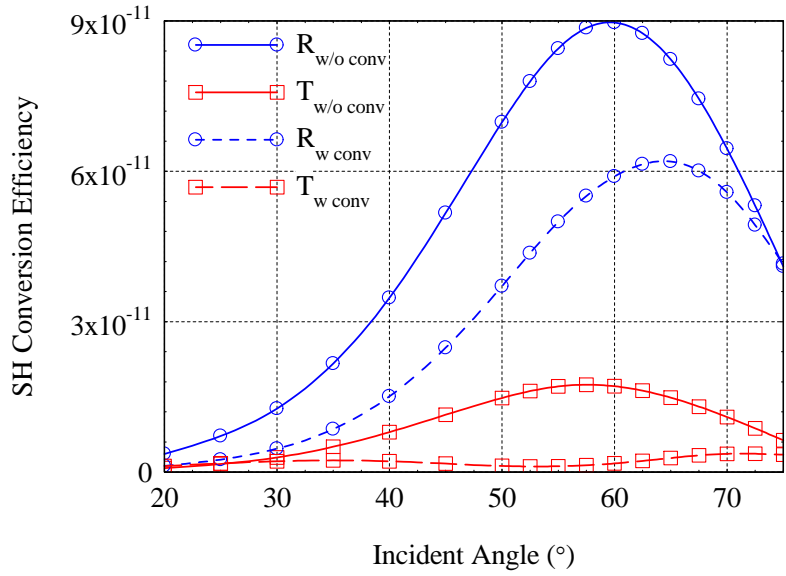

**Fig.5:** TM-incident/TM-detected, reflected and transmitted SH efficiencies, with and without convection. Convection shifts and reduces the amplitude of the reflected signal, while turning the transmission maximum into a minimum. The subscript "w/o" stands for "without", while the subscript "w" stands for "with". We have assumed that $m^*=m_e$, and $n_0=5.8\times10^{22}$/cm$^3$.

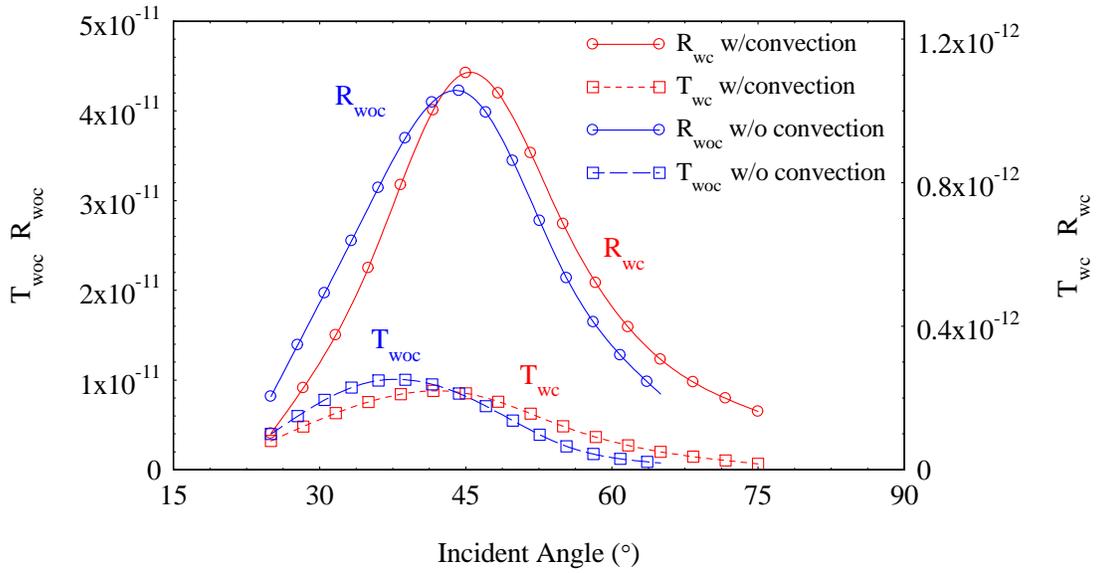

**Fig.6:** TE/TM-incident/TE-detected, reflected and transmitted SH conversion efficiencies, with (right axis) and without (left axis) convection. Convection shifts the peaks to larger angles, but reduces both signals by nearly a factor of 50. The subscript "woc" stands for "without convection", and "wc" stands for "with convection". $m^*=m_e$, and $n_0=5.8 \times 10^{22}/cm^3$.

## EFFECTS FROM BOUND CHARGES

We now change the context and explore the influence of bound charges. In Fig.7 we show SH (400nm) and TH (266nm) emission patterns that originate from a silver nano-pillar having a square cross-section 200nm on each side. The pump field is TM-polarized and is incident from the left. The detected harmonic fields are also TM-polarized. In Figs.7a-b we show SH and TH emission patterns that emerge from the sample if only free charge contributions were turned on (Drude term only in Eq.(1)). In Fig.7c-d we show the corresponding field patterns when both free and bound charges contribute simultaneously. We can see that bound charges can impact field patterns, intensities, and conversion efficiencies. We now recalculate harmonic field emission patterns for a TE-polarized incident pump pulse, otherwise similar in nature to the pulse used in Fig.7, and plot the results in Fig.8. Again we find that the simultaneous presence of free and bound charges can significantly alter the total SH and TH conversion efficiencies and respective field patterns. In all cases adjusting the relative weights of free to bound electron masses and densities can increase bound charge contributions and thus affect the results. Although it may be possible to adjust effective surface and volume coefficients to the extent of obtaining a reasonable fit with experimental results, our simulations suggest that bound charges interfere with free charges and that ignoring their contributions may amount to an important omission in the larger, more complicated dynamical context outlined above.

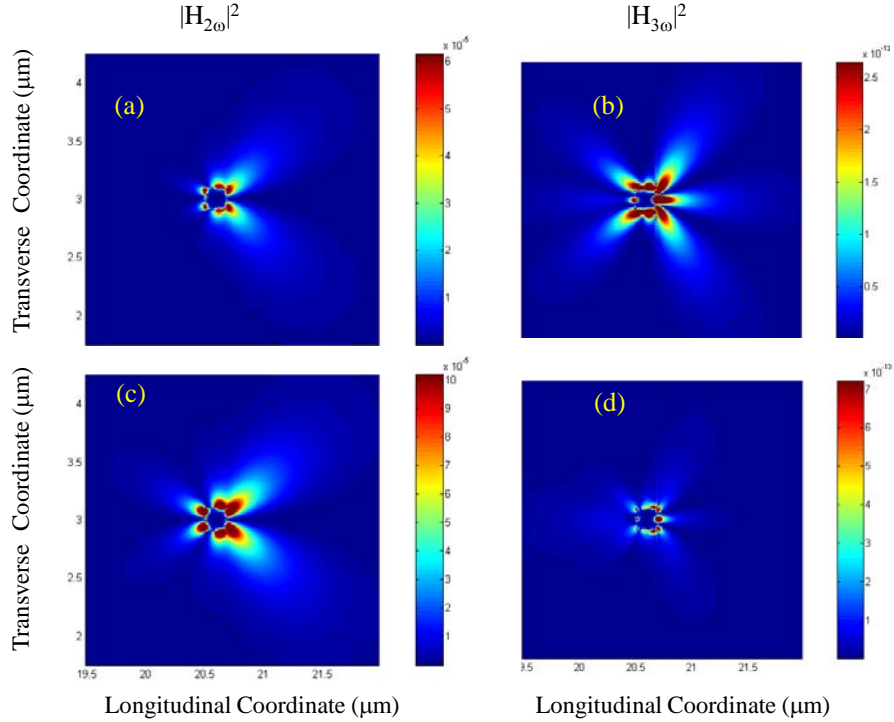

**Fig.7:** A TM-polarized 20fs-pump pulse tuned to 800nm is incident from the left on a silver nano-pillar 200nm on each side. All detected fields are TM-polarized. The snapshots are recorded when the peak of the pump pulse reaches the object. The spatial extension of the pulse is hundreds of times larger than the object, so that this snapshot mimics what one might expect for plane-wave illumination and steady state conditions. (a),(c): TM-polarized SH field patterns for free charges only (a), and free and bound charges simultaneously (c).(b),(d): Same as (a),(c) for the TH fields. Qualitative differences are especially evident for the TH field, where bound charges tend to reinforce the main lobe that points directly backward. In all cases quantitative differences can be read on the side scales. Bound charges enhance SH conversion by a factor of ~2. TH generation is reduced by approximately the same factor. We have assumed $m^*=m_e$, and $n_0=5.8 \times 10^{22}/cm^3$.

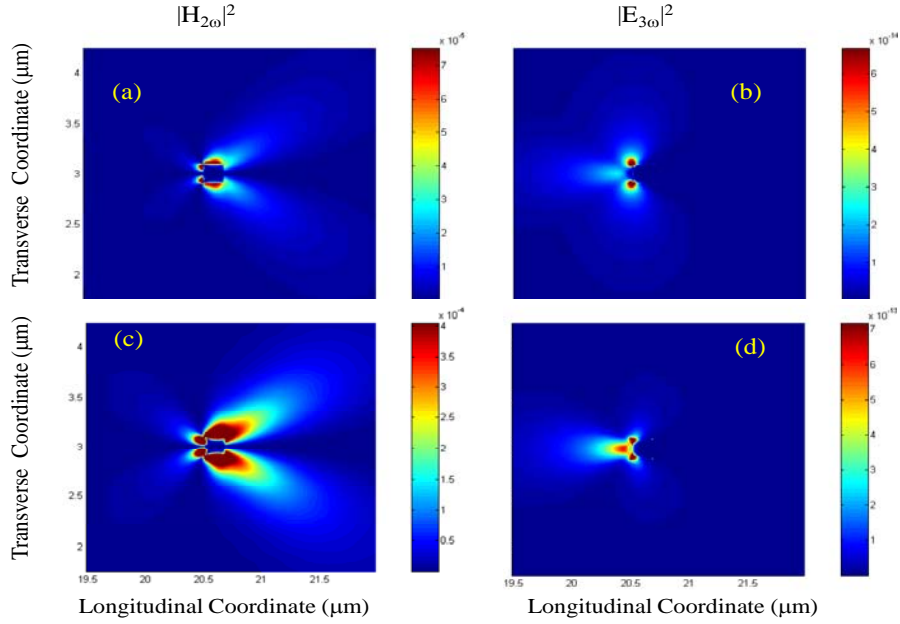

**Fig.8:** Same as Fig.7, except for TE-polarized incident pump pulse. The snapshots are recorded when the peak of the pump pulse reaches the metal object. Unlike Fig.7, the generated SH signals are TM-polarized (H field is shown for convenience), while the TH fields are TE-polarized (E-field is shown for convenience). TM-polarized SH field patterns from free only (a) and from free and bound (c) charges. TE-polarized TH field patterns obtained from free only (b) and free and bound (e) charges. Qualitative differences are visible in both SH and TH patterns. The polarization of a resulting harmonic field under specific pumping conditions may be assessed by a full decomposition of the nonlinear sources in Eqs.(12) and (29).

**ELECTRON GAS PRESSURE**

From Eq.(13) above one may determine that the potential for dynamical modifications of the linear dielectric function and nonlinear contributions to harmonic generation may be significant if the fields become strongly confined to produce either (i) large evanescent wave vectors via strong field localization, or (ii) large spatial derivatives near or just inside the metal surface. In the visible and IR ranges one can always count on the field penetrating inside the metal with consequent rapid spatial modulation. In what follows we concentrate on estimating perturbations to the linear dielectric constant in an infinite array of GaAs-filled, 60nm-wide nanocavities carved on a silver substrate 100nm thick. The substrate is designed to be resonant at $\lambda=1064$nm, and to display enhanced transmission through a combination of plasmonic and Fabry-Perot resonances. The structure is depicted in Fig.9 along with the transmittance for TM-polarized light as a function of center-to-center distance (pitch) between the nanocavities. When the pitch is ~590nm (indicated by the red arrow in Fig.9) the structures displays a transmittance of ~300%, with field amplification values inside the cavity that range from 100 times for the transverse electric field intensity, to approximately 400 times for the magnetic field intensity. These are modest amplifications and correspond to an effective cavity-Q that hovers in the thousands.

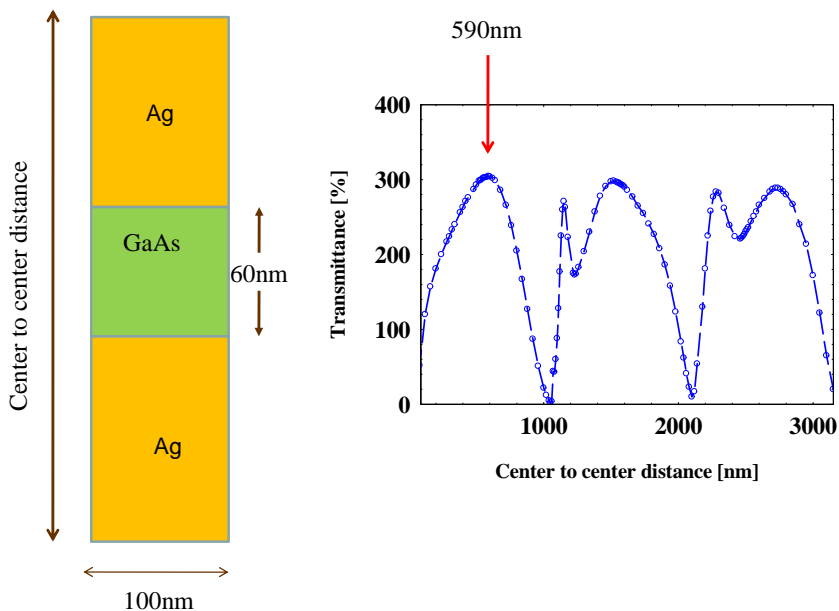

**Fig.9:** Left: GaAs-filled nanocavity 100nm deep and 60nm wide on an infinite array. Departures from these parameters can reduce transmittance and field amplification inside the cavity. Right: Transmittance vs. pitch, normalized to the energy that falls directly on the cavity area. Maxima are Fabry-Perot resonances; minima correspond to Woods anomalies.

The question relative to Eq.(13) and electron gas pressure may be formulated as follows: How large do pressure terms become relative to the linear driving field terms near resonance? We provide a partial answer in Fig.10, where we compare the transverse and longitudinal pressure terms relative to their respective field components, when the peak of an incident 100fs pulse reaches the cavity. The figure shows that the transverse pressure can alter the local, instantaneous transverse forcing term by approximately 1%-2%. Longitudinal pressure changes are strongest at the four corners of the cavity, but amount to a more modest ~0.1%. Of course these rates could increase or decrease depending on the exact magnitude of the Fermi velocity, tuning, and other geometrical factors. Nevertheless, these results show that electron gas pressure may act to either "soften" or "harden" the metal boundaries and to shift resonance conditions. Perhaps more significantly, while it is certainly important to establish the low-intensity bulk metal dielectric constant [68], these results point to the fact that in plasmonic, high-finesse, high-field cavities [56, 67] it may be more important to determine and include terms not needed in more ordinary situations. Our estimates also indicate that in the case of metal layers or metal-dielectric multilayer stacks pressure contributions to the linear dielectric constant amount to at most one part in $10^5$. We will report elsewhere more details on the dynamical effects of electron gas pressure, and the linear and nonlinear optical properties of metallic nanocavities in the enhanced transmission regime. Suffice it to say here that under extreme conditions effects due to electron gas pressure should at a minimum be assessed when strong field confinement occurs at the sub-wavelength scale.

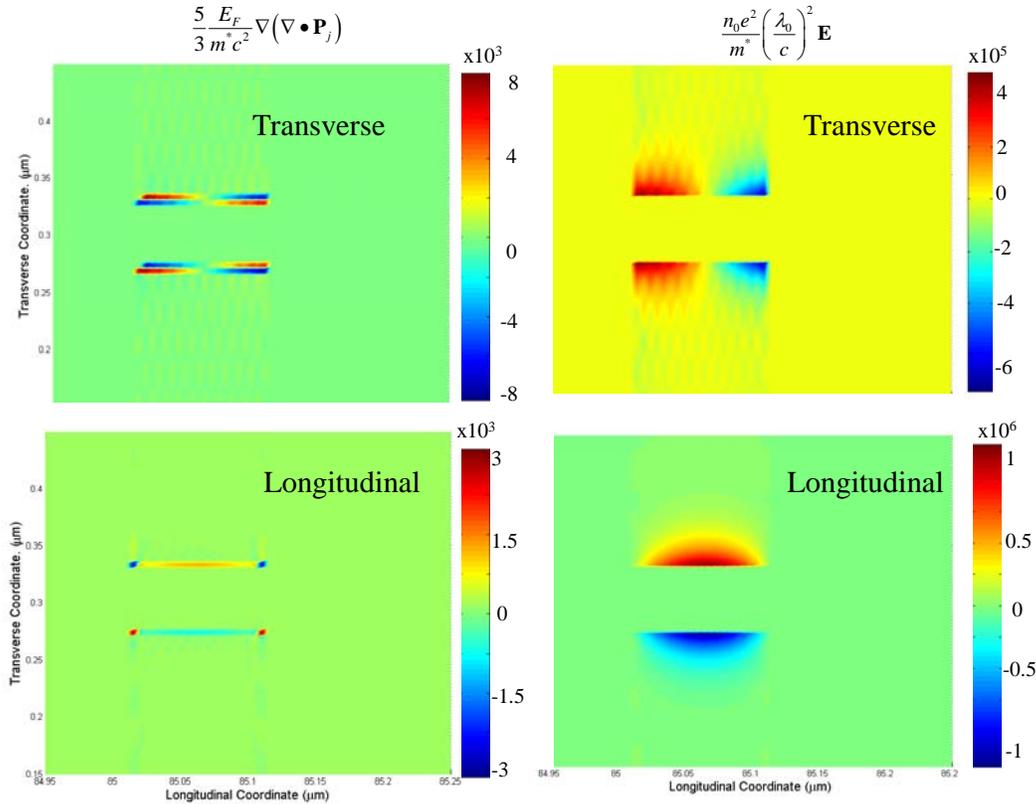

**Fig.10:** Assessing the effects of electron gas pressure. Top (bottom): snapshot of transverse (longitudinal) pressure compared to its driving field term. The scales show that the transverse pressure is spatially modulated with amplitudes of order 1%-2% compared to the driving field term. The longitudinal pressure is modulated along its entire length, but is strongest at the four corners of the cavity. Note: negative pressure simply indicates regions of pressure lower compared to contiguous regions.

## COMPARISON WITH EXPERIMENTAL RESULTS

We now attempt to make a comparison with experimental results, and choose reference [71] because it contains data for a pump field tuned at 1064nm relative to silver for: (i) TE-incident/TM-detected; (ii) TM-incident/TM-detected; (iii) TE-TM incident/TE-detected polarization states. In Fig.11 we show a comparison between our calculations and the data reported in reference [71]. The pulses used in reference [71] were 50 picoseconds in duration, and the silver layer was 400nm thick. Our simulations were carried out using 80fs pulses and a silver layer 150nm thick. As we mentioned earlier, dispersion plays a role only for pulses that are a few optical cycles in duration since the single layer has neither transmission nor reflection features. Substrate thickness becomes unimportant if the layer is substantially opaque. For simplicity, effective mass and density were chosen so as to keep their ratio (i.e. plasma frequency) constant. The input beam used in the experiment was slightly converging (ours was not), with a ~2mm radius at the sample and a peak intensity of ~80MW/cm$^2$. Once these parameters were fixed, Fig.11a was fitted with an incident peak intensity of ~60MW/cm$^2$; in Fig.11b we used ~100MW/cm$^2$ and in Fig.11c ~300MW/cm$^2$. The comparison in Figs.11a-b shows excellent quantitative and qualitative agreement between theory and experiment. Even though Fig.11c also shows excellent qualitative agreement, our input intensity was approximately four times larger than reported in reference [71]. Although these discrepancies are relatively small and could be accounted for by modest power and beam radius fluctuations, more stringent experimental and theoretical tests are underway to further validate the model under a variety of conditions.

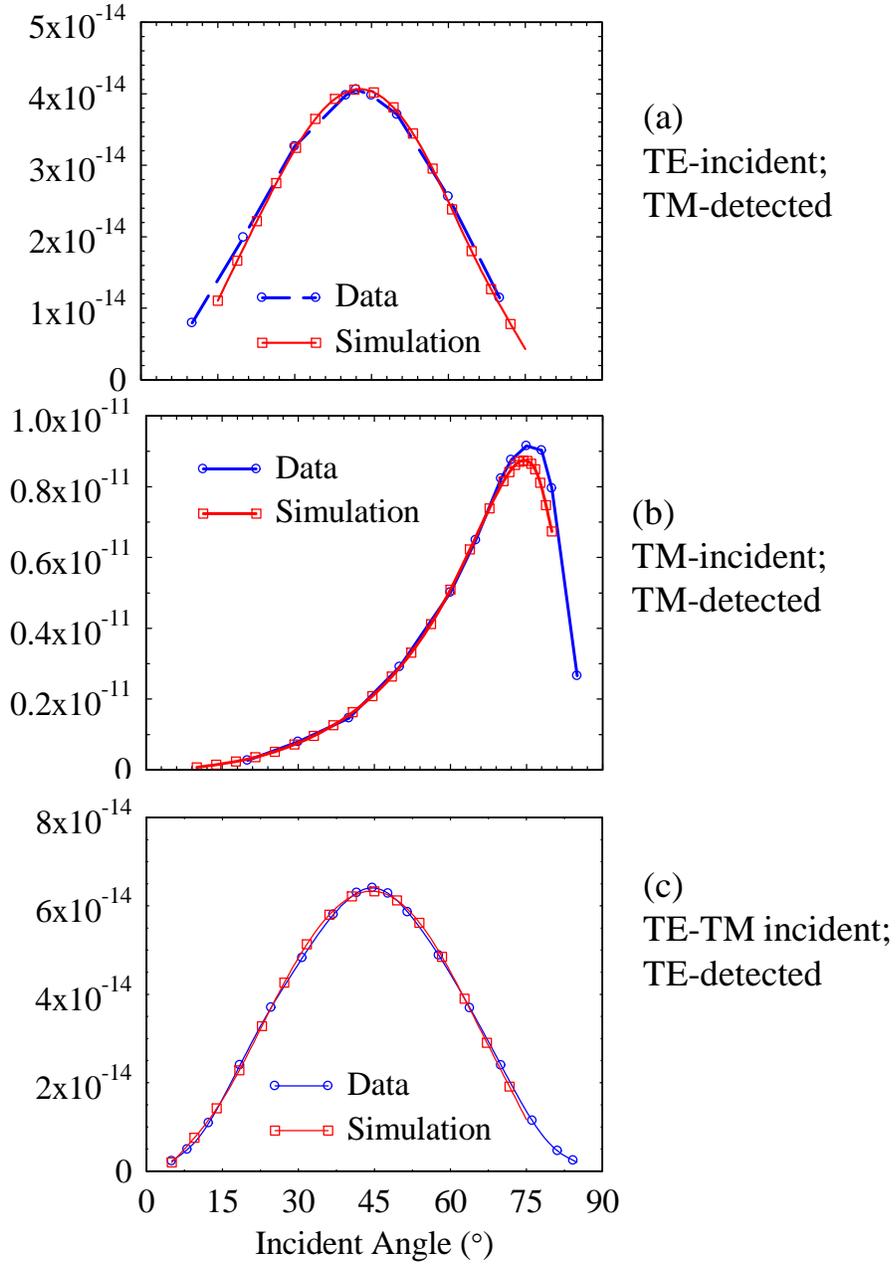

**Fig.11:** SH conversion efficiency under various pumping conditions. The data is taken from reference [71], where the average peak intensity is estimated at ~80MW/cm$^2$. The shapes of theoretical curves agree well with the experimental data in all cases. If we choose $m^*=0.5m_e$, and $n_0=3\times10^{22}$/cm$^3$, for both free and bound charges, excellent quantitative agreement follows with the following choice of incident peak intensities: (a) ~60MW/cm$^2$; (b) ~100MW/cm$^2$; (c) ~300MW/cm$^2$.

**CONCLUSIONS**

We have presented a new model to study propagation phenomena and harmonic generation in nanostructures in the ultrashort pulse regime, with the inclusion of linear and nonlinear effects due to free and bound charges simultaneously. Free electrons are modeled using the hydrodynamics model, comprising effects due to convection and electron gas pressure. Bound charges are modeled by Lorentz oscillators under the action of electric, magnetic and nonlinear restoring forces. We have shown that the influence of bound charges and convection on harmonic generation can be qualitatively and quantitatively appreciable. In addition, we have briefly discussed dynamical changes that can occur in resonant plasmonic nanocavities as a result of electron gas pressure. Although we have derived explicit nonlinear electron gas pressure terms, we have not discussed their influence on harmonic generation. We will do so in a different setting, in the more specific context of GaAs-filled resonant nanocavities. Finally, we found our calculations to be in good qualitative and quantitative agreement with the data of reference [71], where SHG from a silver film was reported under different incident pumping conditions. Although some differences remain and more tests cases are actively under consideration, our calculations and comparisons so far show that this approach may suffice to explain all relevant features of harmonic generation, provided the model includes all crucial dynamical aspects, namely convection, electron gas pressure, and contributions from core electrons.

**Acknowledgment**: We thank the Army Research Lab, London Office, for partial financial support.